\documentclass[a4paper,fleqn,usenatbib]{mnras}

\usepackage[T1]{fontenc}
\usepackage{ae,aecompl}
\usepackage{newtxtext,newtxmath}

\usepackage{graphicx}	
\usepackage{amsmath}	
\usepackage{pdflscape}
\newcommand{\ha}{H$\alpha$~}
\newcommand{\OIII}{[O{\sc{iii}}]}
\newcommand{\OII}{[O{\sc{ii}}]}
\newcommand{\NII}{[N{\sc{ii}}]}
\newcommand{\SII}{[S{\sc{ii}}]}

\title[Metallicity conversions in MaNGA]{Correction to Conversions between gas-phase metallicities in MaNGA}

\author[J.M. Scudder et al.]{
Jillian M. Scudder,$^{1}$\thanks{E-mail: jillian.scudder@oberlin.edu}
Sara L. Ellison,$^{2}$
Loubna El Meddah El Idrissi$^{1}$, 
\newauthor and Henry Poetrodjojo$^{3, 4}$\\
$^{1}$Department of Physics \& Astronomy, Oberlin College, Oberlin, OH, 44074, USA \\
$^{2}$Department of Physics \& Astronomy, University of Victoria, Finnerty Road, Victoria, British Columbia, V8P 1A1, Canada\\
$^{3}$Research School of Astronomy and Astrophysics, The Australian National University, Cotter Road, Weston, ACT 2611, Australia\\
$^{4}$ARC Centre of Excellence for All Sky Astrophysics in 3 Dimensions (ASTRO 3D)\\
}

\date{Accepted XXX. Received YYY; in original form ZZZ}

\pubyear{2021}

\begin{document}
\label{firstpage}
\pagerange{\pageref{firstpage}--\pageref{lastpage}}
\maketitle

\begin{abstract}
We present a brief correction to \citet{Scudder2021} due to an error in the O3N2 based metallicity calibrations presented in that work. Conclusions are unchanged, but metallicity values shift by $\sim$0.03 dex, with polynomials for affected conversions shifted by the same amount. We present updated materials here.
\end{abstract}




We have identified an error in the calculation of O3N2 based calibrations presented in \citet{Scudder2021}, which were metallicities based on \citet{PP04} O3N2, \citet{Marino2013} O3N2, and \citet{Curti2017} O3N2. [O{\sc{iii}}] fluxes were mistakenly multiplied by 1.33, which is required for R$_{23}$ based calibrations but not for the ([O{\sc{iii}}]$\lambda$5007/H$\beta$)/([N{\sc{ii}}]$\lambda$6584/H$\alpha$) line ratio. Correcting these values systematically decreases the raw metallicity values for these three calibrations, and slightly increases the number of overall spaxels with metallicities. A corrected Table 1 with metallicity values is presented here. All three metallicity catalogues (DR15, DR7, and TYPHOON) were identically processed, and have all now been corrected. 
 We have re-run the rest of the work, and find that the scatter around our polynomial fits is functionally unaffected. 

The polynomial fits themselves shift horizontally or vertically when converting from or to an O3N2 based metallicity calibration into a non-O3N2 based calibration, by somewhere between 0.026 and 0.055 dex. The median magnitude of the vertical shifts between polynomials is 0.032 dex.  Conversions between O3N2 based calibrations and other O3N2 based calibrations are unaffected. We show a sample figure in Figure \ref{fig:polyshift}. We have updated the polynomials presented in Appendix table \ref{tab:smc_values}, and in the full tables presented in the supplementary material. 

Figure 7 of \citet{Scudder2021} is the most directly impacted figure; qualitatively it is virtually the same, as all three populations presented in that Figure were affected by the same systematic error, and for completeness we reproduce it here in Figure \ref{fig:z94_poly}.  The right hand panel of Figure 8 of \citet{Scudder2021} is the only figure that has a visible change with the update of these metallicities, with the reduction of offsets between polynomials for PP04 O3N2 based metallicities into any other metallicities reduced by 0.04 dex to 0.1 dex. We thus show it here as Figure \ref{fig:superfig_polyoffset}. The median offset across all polynomials is only reduced by 0.003 dex relative to that reported in \citet{Scudder2021}. 

All remaining figures in \citet{Scudder2021} are affected by $\lessapprox$ 0.003 dex, with the updated typically reducing scatter, and are not visibly different from those published. Values in text are either identical or correct within 0.003 dex. Tables not reproduced here are also completely unchanged from the original published version. 
Supplementary online materials (all versions of Figures 2, 3, 4, 7 in \citealt{Scudder2021}, and the full tables of polynomials) have been fully updated.  

\begin{table}
	\centering
	\caption{Total number of metallicity values per calibration, for both the SMC and MW dust correction models, after S/N cuts, BPT classifications, and including an \ha EW cut.}
	\label{tab:oh_nums}
	\begin{tabular}{lcr}
	\hline
Calibration & SMC spaxels & MW spaxels\\ 
\hline 
Z94 & 1,047,232 & 858,244 \\ 
M91 & 1,060,728  &953,147  \\ 
KK04 & 1,065,094 & 989,498  \\ 
KE08 & 1,069,318  & 1,027,191 \\ 
D16 & 1,012,424  & 890,787 \\ 
PP04 N2 & 1,091,383 & 1,055,164 \\ 
M13 N2 & 1,092,348 &1,055,271 \\ 
C17 N2 & 1,081,521 & 1,051,017 \\ 
PP04 O3N2 & 1,091,786  & 1,054,508 \\ 
M13 O3N2 & 1,087,266  & 1,048,824\\ 
C17 O3N2 & 1,092,355  & 1,055,359 \\ 
		\hline
	\end{tabular}
\end{table}

\begin{table}
	\centering
	\caption{In ascending typical 2$\sigma$ scatter, we present the emission line permutations between calibrations. For each set of calibrations which match the inclusions/exclusions, we find the typical offset of the 2$\sigma$ contour (the median absolute value of the 2$\sigma$ residuals) from our polynomial fit. We also include the smallest and largest $2\sigma$ residuals for each set. The median (and range) of the 2$\sigma$ scatter is smallest for all calibrations which have full overlap in their emission line requirements: the top row includes all of the O3N2-based calibrations. }
	\label{tab:eline_sort}
\begin{tabular}{r|lccc}
\hline
Lines excluded & Lines in overlap & \multicolumn{3}{|c||}{2$\sigma$ scatter (dex)} \\ 
($x$ but not $y$) & ($x$ \& $y$) & $\mu_{1/2}$ & min  & max  \\ 
\hline 
Full overlap & \OIII, H$\alpha$, \NII, H$\beta$ & 0.0025 & 0.0022 & 0.0036\\ 
Full overlap & H$\alpha$, \NII & 0.0525 & 0.002 & 0.1549\\ 
Full overlap & \OII, \OIII, \NII, H$\beta$ & 0.0602 & 0.0135 & 0.1412\\ 
H$\beta$, \OIII & H$\alpha$, \NII & 0.0613 & 0.0477 & 0.0919\\ 
H$\beta$, \SII & H$\alpha$, \NII & 0.0697 & 0.0522 & 0.0895\\ 
\SII & H$\alpha$, H$\beta$, \NII & 0.0792 & 0.0712 & 0.1127\\ 
\OII & H$\beta$, \NII, \OIII & 0.0897 & 0.0615 & 0.1391\\ 
H$\beta$, \OII, \OIII & \NII & 0.1124 & 0.0674 & 0.1767\\ 
H$\alpha$ & H$\beta$, \NII, \OIII & 0.144 & 0.108 & 0.1599\\ 
H$\alpha$, \SII & H$\beta$, \NII & 0.1564 & 0.1358 & 0.1863\\
H$\alpha$ & \NII & 0.1724 & 0.117 & 0.2023\\ 
\OIII & H$\alpha$, H$\beta$, \NII & 0.1907 & 0.1892 & 0.1997\\ 
\OII, \OIII & H$\beta$, \NII & 0.2687 & 0.1802 & 0.2838\\ 
\hline
\end{tabular}
\end{table}

%
  
  \begin{figure}
  \includegraphics[width=0.8\columnwidth]{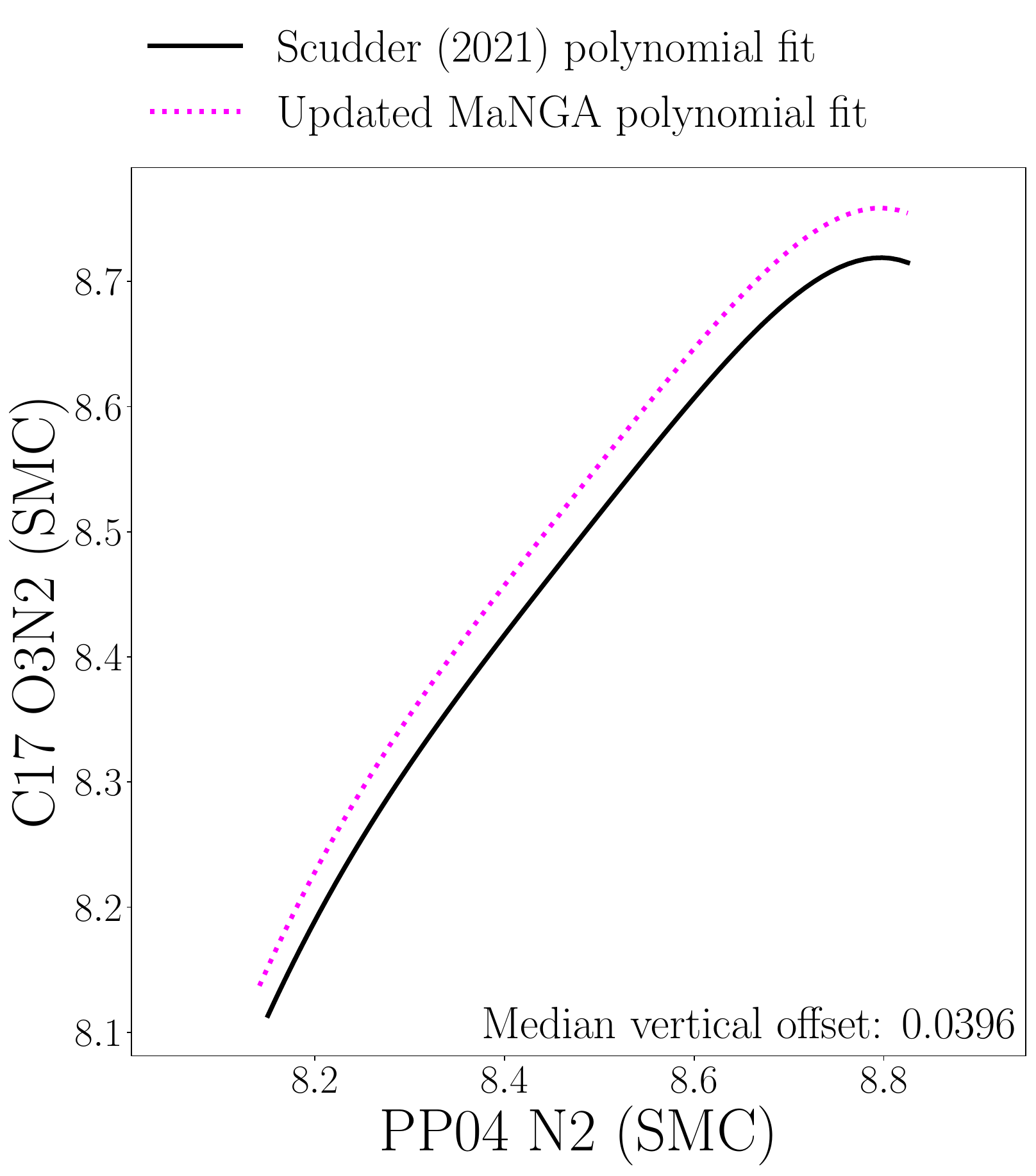}
  \caption{Comparison of a polynomial lines of best fit as published in \citet{Scudder2021} in a solid black line, and the corrected metallicities in a pink dotted line. The median vertical offset between polynomials for the range in x values with polynomial coverage is plotted in the lower left corner. In this case, the difference between polynomials is about 0.04 dex. This figure is representative of the change in the polynomials.}
  \label{fig:polyshift}
\end{figure}

\begin{figure*}
  \includegraphics[width='7in]{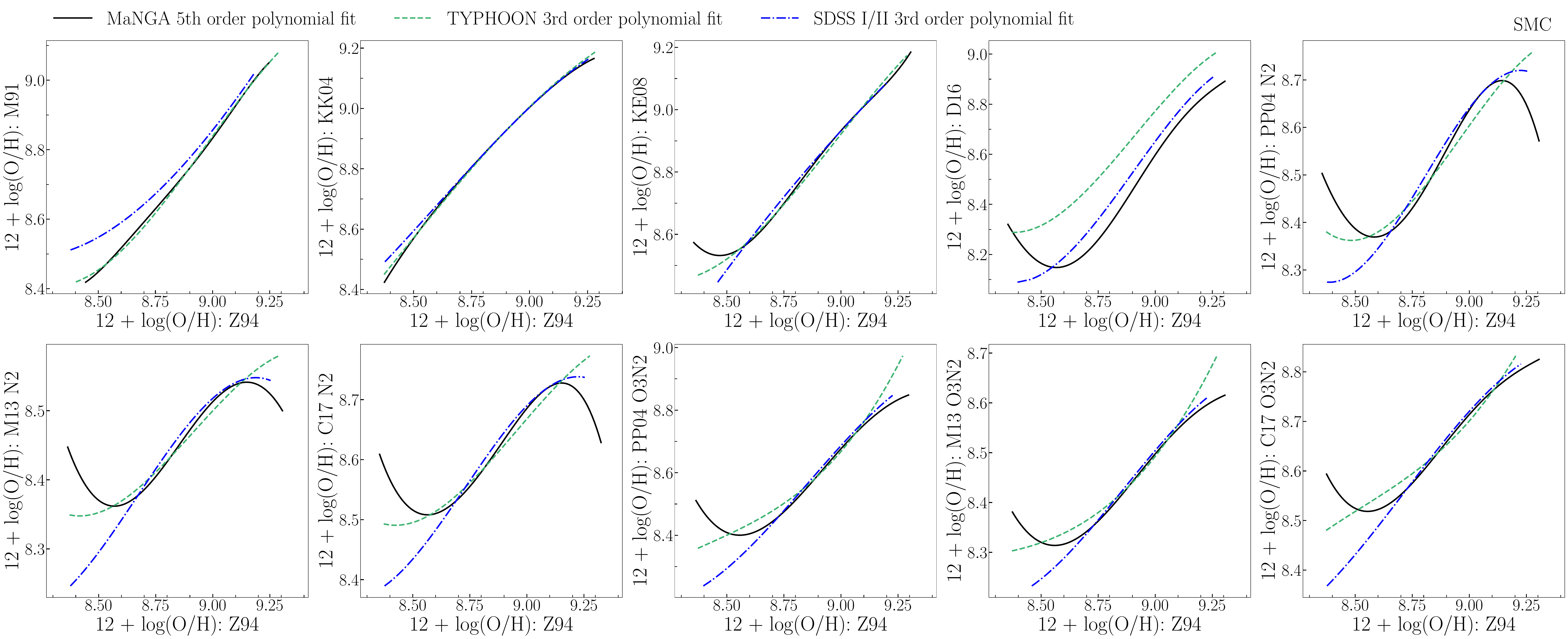}
  \caption{Comparison of the polynomial lines of best fit. The 5th order polynomial fit to the MaNGA data presented here are plotted in a black solid line. We plot the 3rd order polynomial fits to the DR7, which are also affected by this metallicity erratum, in a blue dot-dashed line. 3rd order polynomial fits to the TYPHOON data, also recalculated here, are plotted in a dashed green line.}
  \label{fig:z94_poly}
\end{figure*}

\begin{figure*}
  \includegraphics[width=.495\textwidth]{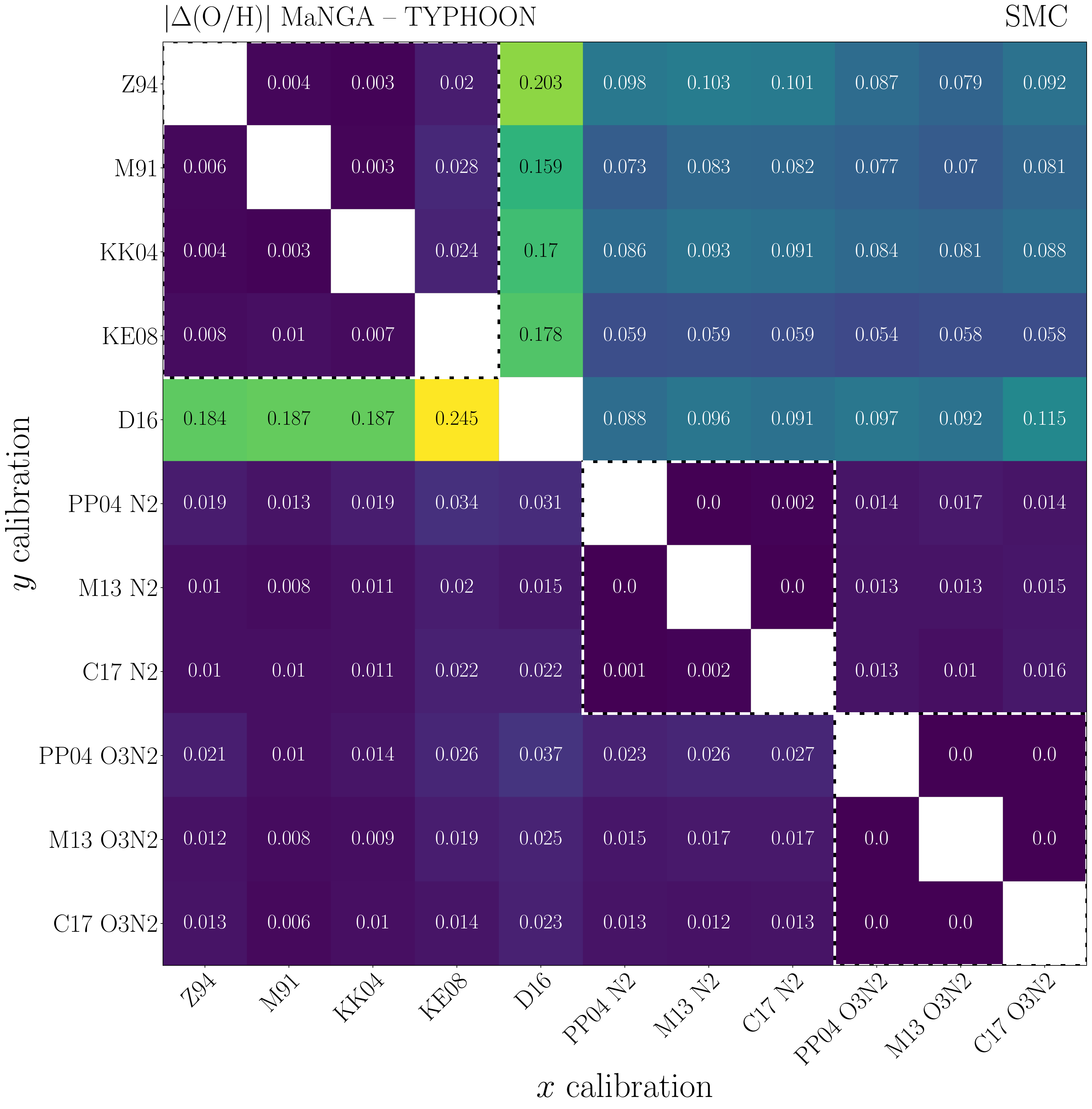}
  \includegraphics[width=.495\textwidth]{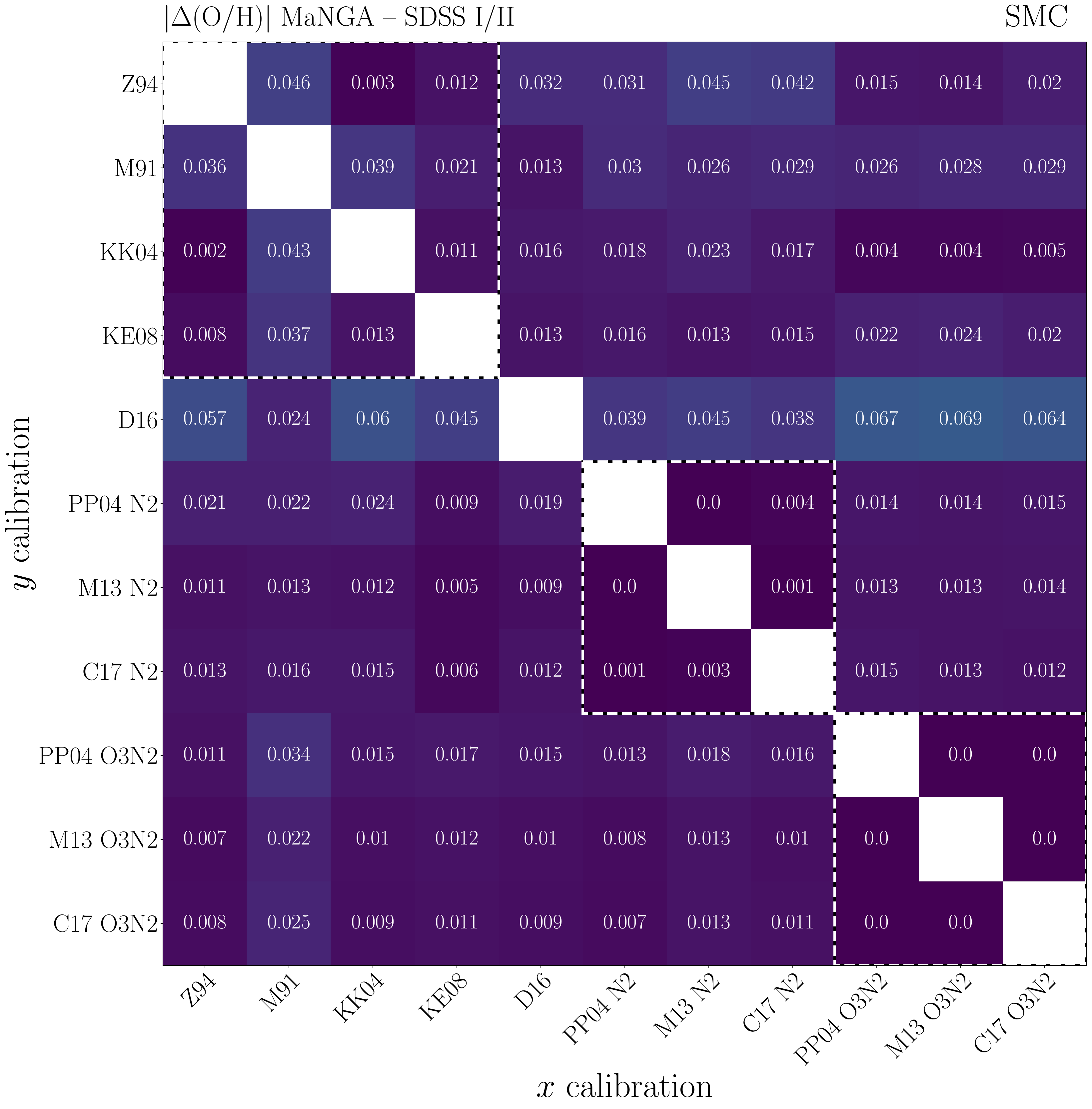}
  \caption{Comparison of the differences between polynomial lines of best fit. The left panel is functionally unchanged, with median offsets still at 0.019 dex. The median difference between MaNGA \& DR7 (right) is reduced by 0.003 dex to 0.014 dex, with the published trend of PP04 O3N2 metallicities being slightly more offset now removed.}
\label{fig:superfig_polyoffset}
\end{figure*}

\section*{Data Availability}
The emission line data underlying \citet{Scudder2021} are publicly available as part of the MaNGA DR17 data release, available at \href{https://www.sdss.org/dr17/}{https://www.sdss.org/dr17/}. Metallicity values themselves are avail- able upon reasonable request to the corresponding author.




\bibliographystyle{mnras}
\bibliography{new_library} 

\begin{thebibliography}{}
\makeatletter
\relax
\def\mn@urlcharsother{\let\do\@makeother \do\$\do\&\do\#\do\^\do\_\do\%\do\~}
\def\mn@doi{\begingroup\mn@urlcharsother \@ifnextchar [ {\mn@doi@}
  {\mn@doi@[]}}
\def\mn@doi@[#1]#2{\def\@tempa{#1}\ifx\@tempa\@empty \href
  {http://dx.doi.org/#2} {doi:#2}\else \href {http://dx.doi.org/#2} {#1}\fi
  \endgroup}
\def\mn@eprint#1#2{\mn@eprint@#1:#2::\@nil}
\def\mn@eprint@arXiv#1{\href {http://arxiv.org/abs/#1} {{\tt arXiv:#1}}}
\def\mn@eprint@dblp#1{\href {http://dblp.uni-trier.de/rec/bibtex/#1.xml}
  {dblp:#1}}
\def\mn@eprint@#1:#2:#3:#4\@nil{\def\@tempa {#1}\def\@tempb {#2}\def\@tempc
  {#3}\ifx \@tempc \@empty \let \@tempc \@tempb \let \@tempb \@tempa \fi \ifx
  \@tempb \@empty \def\@tempb {arXiv}\fi \@ifundefined
  {mn@eprint@\@tempb}{\@tempb:\@tempc}{\expandafter \expandafter \csname
  mn@eprint@\@tempb\endcsname \expandafter{\@tempc}}}

\bibitem[\protect\citeauthoryear{Curti, Cresci, Mannucci, Marconi, Maiolino  \&
  Esposito}{Curti et~al.}{2017}]{Curti2017}
Curti M.,  Cresci G.,  Mannucci F.,  Marconi A.,  Maiolino R.,   Esposito S.,
  2017, \mn@doi [MNRAS]
  {10.1093/mnras/stw2766}, 465, 1384

\bibitem[\protect\citeauthoryear{Marino et~al.,}{Marino
  et~al.}{2013}]{Marino2013}
Marino R.~A.,  et~al., 2013, \mn@doi [A\&A]
  {10.1051/0004-6361/201321956}, 559, A114

\bibitem[\protect\citeauthoryear{Pettini \& Pagel}{Pettini \&
  Pagel}{2004}]{PP04}
Pettini M.,  Pagel B. E.~J.,  2004, \mn@doi [MNRAS] {10.1111/j.1365-2966.2004.07591.x}, 348, L59

\bibitem[\protect\citeauthoryear{Scudder, Ellison, Idrissi  \&
  Poetrodjojo}{Scudder et~al.}{2021}]{Scudder2021}
Scudder J.~M.,  Ellison S.~L.,  Idrissi L. E. M.~E.,   Poetrodjojo H.,  2021,
  \mn@doi [MNRAS]
  {10.1093/mnras/stab2339}, 507, 2468

\makeatother
\end{thebibliography}


\appendix

\section{Tables}
In this Appendix, we provide a sample few rows of the tables which list conversions between all calibrations, the number of spaxels used in the fitting procedure, the range of validity, and the polynomial fits used in this work as an example of the data structure. 
\begin{landscape}
\begin{table}
	\centering
	\caption{Summary of the median offset from a 5th order polynomial best fit in both positive and negative directions, for the contour that encloses 95.5 per cent of the data. For each metallicity calibration pairing, the number of spaxels which are present is also recorded, for the SMC dust correction curve. The full table, along with the same for all other pairings, and for the MW dust correction curve, is available as supplementary material.\label{tab:smc_values}}
\begin{tabular}{lccccccccccll}
\hline
Conversion $x \rightarrow y$ & $n$ spaxels & 2$\sigma$ scatter &  Range of validity & \multicolumn{6}{|c|}{Polynomial fit: a + b$x$ + c$x^2$ + d$x^3$ + e$x^4$ + f$x^5$} \\ 
&&& $\left(12+\text{log(O/H)} \right)$ & a & b & c & d & e & f \\ 
\hline 
 Z94 $\rightarrow$ PP04 O3N2 & 1047174 & [-0.1235, 0.1251] & [8.3647, 9.2955] & -8255.57111302 & + 5500.47948047 &-1412.33588325 & + 176.61349745 &-10.82276874 & + 0.26107276 \\ 
\hline 
 Z94 $\rightarrow$ M13 O3N2 & 1046104 & [-0.0826, 0.0836] & [8.3745, 9.3053] & -21025.73022481 & + 12464.89658234 &-2935.1585598 & + 343.53656396 &-19.99737841 & + 0.46338882 \\ 
\hline 
 Z94 $\rightarrow$ C17 O3N2 & 1047150 & [-0.0891, 0.085] & [8.3745, 9.3051] & -30018.98525082 & + 17956.62194749 &-4268.0189955 & + 504.38192063 &-29.65474639 & + 0.69430492 \\ 
\hline 
 M91 $\rightarrow$ Z94 & 1037616 & [-0.0264, 0.0206] & [8.4035, 9.0696] & -82822.51326969 & + 51125.46960079 &-12549.66580389 & + 1532.47593546 &-93.14893094 & + 2.25576348 \\ 
\hline 
 M91 $\rightarrow$ KK04 & 1055420 & [-0.0199, 0.0098] & [8.4035, 9.0712] &  + 49862.66651585 &-26473.36992983 & + 5590.08956813 &-586.27362519 & + 30.50799985 &-0.62930761 \\ 
 &  & \textit{lower branch:} & [7.8508, 8.2351] & -25.16773802 & + 7.44906684 &-0.40975953 &&&\\ 
 \hline 
...&...& ...& ... &...&...& ...& ... &...&...\\
 \hline 
 \end{tabular}
\end{table}
\end{landscape}

\bsp	
\label{lastpage}
\end{document}